\documentclass[aps,twocolumn,showpacs,showkeys,floatfix]{revtex4}
\usepackage{graphicx}% Include figure files
\usepackage{dcolumn}% Align table columns on decimal point
\usepackage{bm}% bold math

\begin{document}

%\preprint{Manuscript ??????}

\title{Explanation for Anomalous Shock Temperatures Measured by 
   Neutron Resonance Spectroscopy}

%\date{February 27, 2007}
%\date{March 12, 2007 -- LA-UR-07-1981}
%\date{April 18, 2007 -- LA-UR-07-1981}
\date{June 29, 2007 -- LA-UR-07-1981}

\author{Damian C. Swift}
\email{dswift@lanl.gov}
%\homepage{http://public.lanl.gov/dswift}
\author{Achim Seifter}
\author{David B. Holtkamp}
\author{Vincent W. Yuan}
\author{David Bowman}
\author{David A. Clark}
%\author{William T. Buttler}
\affiliation{%
   Los Alamos National Laboratory,
   Los Alamos, New Mexico 87545, USA
}

\begin{abstract}
Neutron resonance spectrometry (NRS) has been used to measure the temperature
inside Mo samples during shock loading.
The temperatures obtained were significantly higher than predicted assuming
ideal hydrodynamic loading.
The effect of plastic flow and non-ideal projectile behavior were assessed.
Plastic flow was calculated self-consistently with the shock jump conditions:
this is necessary for a rigorous estimate of the locus of shock states
accessible.
Plastic flow was estimated to contribute a temperature rise of 53\,K
compared with hydrodynamic flow.
Simulations were performed of the operation of the explosively-driven projectile
system used to induce the shock in the Mo sample.
The simulations predicted that the projectile was significantly curved on impact,
and still accelerating.
The resulting spatial variations in load, including radial components of 
velocity, were predicted to increase the apparent temperature that would be
deduced from the width of the neutron resonance by 160\,K.
These corrections are sufficient to reconcile the apparent temperatures 
deduced using NRS with the accepted properties
of Mo, in particular its equation of state.
\end{abstract}

% 06.60.Jn High-speed techniques (microsecond to femtosecond)
% 07.35.+k High-pressure apparatus; shock tubes; diamond anvil cells
% 62.50.+p High-pressure and shock wave effects in solids and liquids
%% 25.40.Ny  Resonance reactions
% 29.30.Hs Neutron spectroscopy
%% 64.30.+t EOS for specific substances
\pacs{06.60.Jn, 07.35.+k, 62.50.+p, 29.30.Hs}
%\keywords{Suggested keywords}%Use showkeys class option for keyword display
\keywords{shock physics, temperature measurement, neutron resonance spectroscopy}

\maketitle

%\section{Introduction}
A long-standing problem with shock wave experiments on 
condensed matter is the difficulty of measuring
the temperature of the shocked state before it is destroyed by
release waves.
Most measurements have been made using photon emission spectroscopy
(pyrometry), but many substances of interest to science and engineering 
(e.g. metals) are opaque in the relevant 
region of the spectrum: infra-red through visible for shocks up to
the terapascal regime.
Emission from an opaque material comes from material within the photon
skin depth of the surface, which
generally cannot be maintained at the pressure of the initial shock for long
enough to allow useful emission spectra to be collected.
A transparent window can be placed in contact with the sample to maintain
an elevated pressure at the surface until released from the free surface of the 
window or the projectile.
However, the mismatch in shock impedance must be
taken into account, along with the effect of heat conduction.

Neutron resonance spectroscopy (NRS) has been investigated as a
fundamentally different technique for measuring the
temperature inside a dynamically-loaded specimen,
irrespective of its photon opacity
\cite{Yuan05}.
Nuclear resonances are characterized by the energy and line width
in the rest frame of the nucleus, i.e. with respect to the speed of
incoming neutrons relative to the nucleus.
The resonance is manifested as the variation of attenuation with
neutron energy.
In an NRS measurement of a shocked sample,
neutrons of a range of energies interact with a volume of material.
The resonance measurement is a convolution of the resonance in the rest frame
of a nucleus with the velocity distribution of the nuclei in the sample volume,
which depends on the sample temperature.
A pulse of neutrons of a range of energies passes through the specimen,
chosen to have a measurable nuclear resonance. 
The pulse has a finite duration $o(200)$\,ns and the spectrum varies
with time, lower energy neutrons arriving later.
The spectrum of neutrons is measured after passing through the
specimen; the temperature can be inferred from the width of the
resonance.
The resonance is also shifted in energy by the relative speed of the
specimen with respect to the neutrons, so the spectrum can also provide a
measurement of material velocity inside the sample -- a net average speed
which shifts the centroid of the resonance.
In contrast, almost all velocity measurements are made optically
(usually by the Doppler shift) at the surface of the specimen.

To perform a single-shot NRS measurement of a shocked state,
it was necessary to collect a statistically significant neutron spectrum
within the time for which the sample was in the shocked state,
which is of the order of 1\,$\mu$s for well-understood shocks as generated
by the impact of disk-shaped projectiles \cite{Bushman93}.
To achieve the necessary neutron intensity, it was necessary to design
a dedicated $^{238}$U spallation target and moderator, and to induce 
spallation neutrons using a pulse of 800\,MeV protons which had been
accumulated in the proton storage ring (PSR) of Los Alamos' LANSCE accelerator.

In any shock loading experiment, a difficulty is always the synchronization
of measurements with the shock event.
It is particularly challenging to synchronize an impact event, which has a long 
delay during acceleration and coasting of the projectile,
with a diagnostic pulse from a particle accelerator, which is generally
not designed to be triggered externally with a short latency or high
{\it a priori} precision in delivery of the diagnostic pulse.
For the trial NRS experiments, the projectile was accelerated by
detonating solid chemical explosive,
using the `Forest Flyer' design which 
gave an uncertainty in impact time from the trigger signal of $o(100)$\,ns.
The Forest Flyer produced a non-isotropic distribution of high speed fragments.
In order to protect the LANSCE beamline from damage and breach of vacuum,
the shock experiment was tilted so that
the projectile and shock state in the sample were inclined at
$55^\circ$ with respect to the neutrons.
The inclination was important in interpreting the NRS measurements,
as discussed below.
The separation between the projectile and the target was 20\,mm,
not 15\,mm as reported before \cite{Yuan05}.
%accurate specifications matter in dynamic loading experiments
%and were a recurring problem for the shock loading system.

NRS experiments have been performed on the
reaction products of detonating chemical explosive, and on Mo as it is a
standard reference material for high pressure work \cite{Yuan05}.
The temperature inferred was significantly higher than expected from shock
calculations using the best available
equations of state for Mo \cite{Yuan05}.
The Mo was doped with 1.7\,at.\%\ of $^{182}$W, and
the projectile was accelerated by high explosive.
NRS temperature measurements were made on two nominally
identical shock experiments.
If the shock states were identical, these measurements
could be combined with a root-mean-square uncertainty
(product of the probability distribution function for each measurement),
as was done previously: $883\pm 46$\,K.
The separate measurements differed by around the sum of their neutron-counting
uncertainties, which is not inconsistent with an identical temperature.
Explosively-driven experiments often exhibit some round-to-round variability.
Variations in surface velocity of $O(10\%)$ were observed in similar
experiments \cite{Swift_pyro_07}, and variations in the timing of the
neutrons with respect to the shock may lead to variations in NRS temperature,
as discussed below.
Round-to-round variability argues for combining the two measurements as the
sum of their probability distribution functions: $872\pm 90$\,K.
Non-idealities in the loading are more likely
to cause an increase than a decrease in the apparent temperature,
so the lower measured temperature may be more accurate.
The sensitivity to temperature on the principal shock Hugoniot of Mo 
to uncertainties in the equation of state (EOS)
is believed to be too small to account for the
observed discrepancy \cite{Greeff05}.
Here we explain the temperature discrepancy by taking account of plastic
heating in the sample, and by considering details of the shock loading
system used, which induced perturbations in the shock state that affect the
apparent material temperature as measured by NRS.

%\section{Plastic flow}
In an ideal planar impact experiment, the strain applied to the sample 
is uniaxial.
Uniaxial strain applied to a solid induces shear stresses,
and for Mo at the $\sim$60\,GPa pressures of the NRS experiments
the shear stresses induce plastic flow.
Compared to shock temperatures in a material without shear strength,
plastic flow causes additional heating.
The previous NRS temperatures \cite{Yuan05} were calculated from the scalar EOS,
and neglected the effects of plastic heating.
%, incorrectly assessing them
%as small for Mo.

\begin{figure}
\begin{center}\includegraphics[scale=0.7]{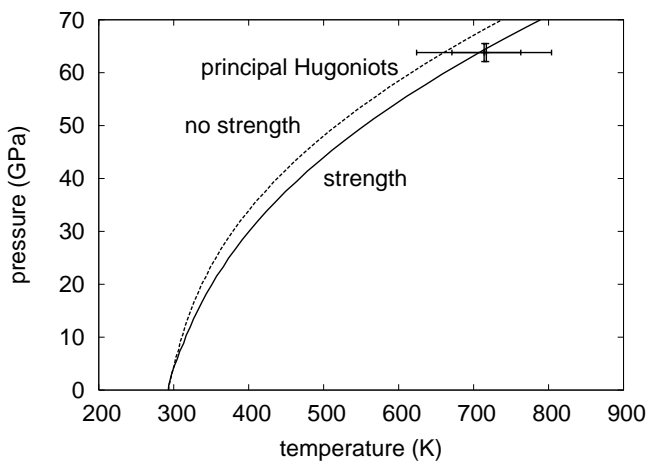}\end{center}
\caption{Corrected temperature inferred from neutron resonance data,
   compared with the shock Hugoniot for Mo with and without strength.
   The smaller error bar is the product distribution
   (identical state preparation);
   the larger is the sum distribution (variation in states).}
\label{fig:nrs_tp}
\end{figure}

The shock Hugoniot of Mo was calculated with and without the contribution from
plastic work.
Plastic work increases the thermal contribution
to the EOS, generally increasing the pressure for a given
compression.
The effect of material strength, i.e. of elastic stress and plastic flow,
was treated self-consistently in a numerical solution of the
Rankine-Hugoniot equations for shock compression \cite{RH,Swift_gen0d_07}.
This is necessary for a rigorous prediction of heating.
Material strength was treated using the
Steinberg-Guinan model \cite{Steinberg80,Steinberg96}.
The effect of plastic work was calculated to be around 53\,K at shock
pressures around 63\,GPa
\cite{Swift_pyro_07}
-- a significant contribution, but not enough to reconcile the
temperature discrepancy (Fig.~\ref{fig:nrs_tp}).
This analysis depends on the accuracy of the Steinberg-Guinan model at
these pressures on the Hugoniot.
Supporting evidence is provided by surface Doppler velocimetry measurements
made of these experiments: the onset of release from the peak velocity
is marked by an elastic release wave of amplitude consistent with the
flow stress predicted using the Steinberg-Guinan model
\cite{Swift_pyro_07}.
The magnitude of plastic heating may be estimated simply,
by multiplying the elastic component of stress by the change in volume
as the sample is compressed.
At 63\,GPa, the compression on the principal shock Hugoniot
of Mo is 0.85, which is almost entirely plastic.
Mo exhibits work-hardening, pressure-hardening, and thermal softening,
so the integrated plastic work depends on the precise deformation history.
However, Steinberg-Guinan flow stresses $Y$ for Mo generally fall in the range
1.6-2.8\,GPa, following the convention that the elastic contribution
to the normal stress is $\frac 23 Y$.
Thus, without accounting for the precise deformation path, plastic heating
should be approximately 60-100\,K.

Pyrometry measurements have been made of the temperature of Mo 
on release from shocking to similar pressures, with release into
a LiF window ($\sim$25\,GPa residual pressure) and into vacuum.
These temperatures were also higher than predicted without accounting for
plastic heating.
Pyrometry is prone to other systematic errors, such as thermal emission from
the shocked window or from gas or glue compressed in the gap between the
sample and the window, and enhanced plastic heating from the deformation of
surface features such as machining marks.
The total power in thermal emission varies with the fourth power of temperature,
and pyrometry measurements are often more accurate at shorter wavelengths
where the power varies with higher powers of temperature.
Pyrometry is therefore prone to inaccuracy from spatial or temporal variations
in temperature, whereas NRS measures the average temperature.
However, the inclusion of plastic heating in Mo brought the predicted surface 
temperatures into reasonable agreement with the pyrometry data
\cite{Swift_pyro_07}.

The speed of the projectile was not measured in the NRS or pyrometry
experiments, so the shock state could not be inferred directly from the
published Hugoniot data for Al and Mo.
The shock pressure was inferred from the peak free surface velocity
observed in each experiment, using mechanical EOS $p(\rho,e)$ derived
from the Hugoniot data \cite{Steinberg96}.
For a given projectile speed or for a given free surface velocity,
the shock pressure depends on the shear modulus and flow
stress assumed in the projectile and sample.
Thus the `experimental' pressures as well as the predicted temperatures 
were adjusted when strength was included.

%\section{Non-ideality of the projectile}
The pressures induced by the detonation of the chemical explosive
in the Forest Flyer were much higher than the flow stress of the
Al projectile. According to continuum dynamics simulations,
the design used for the Mo NRS experiments
suffered from hydrodynamic features which deform and damage the projectile
\cite{Swift_heflyer_07}.
In particular, the case profile
is likely to produce curvature of the projectile.
The projectile would also be accelerating on impact.
If the projectile is still accelerating, there must be a gradient of pressure
and compression through it.
On impact, these gradients induce a shock wave with a driving pressure
which increases with time, leading to
an increasing particle velocity.
Unless this variation is taken into account, 
the broader peak could be attributed to a higher material temperature.
If the projectile is not flat, and particularly if there are radial variations in 
its speed at the time of impact, there will be radial variations in
pressure, temperature, compression, and particle speed.
Any radial component of particle velocity gives a different relative velocity
compared with the neutrons, going around the azimuth.
This variation in relative speed broadens the neutron attenuation peak,
which could again be attributed to a higher material temperature unless
taken into account.
Qualifying experiments were performed on the Forest Flyer system as used
in the NRS experiments, but they were based on arrival time measurements
and did not probe the detailed shape or density distribution of the
projectile.
Proton radiographs were subsequently obtained of a similar Forest Flyer design,
initiated with a plane-wave lens, 
showing curvature of the projectile in close agreement with the
simulations in the region of the projectile affecting the W-doped Mo
\cite{Swift_heflyer_07}.
In the NRS experiments, 
the explosive charge was initiated by 61 detonators fired simultaneously.
Detonator misfires may occur, which could lead to
additional distortion of the projectile and a higher apparent temperature;
misfires are a possible explanation for the two different NRS temperatures.

%\subsection{Simulated neutron resonance spectra}
The simulations were used to predict the variation of
compression, temperature, and particle velocity in the doped Mo
as a function of time.
In principle, the time-dependent fields could be used to simulate
the neutron attenuation with the time-dependent neutron spectra.
A slightly simpler procedure was adopted here, predicting the neutron
attenuation as a function of energy at a series of instants in time.
%This procedure neglects the contribution of temporal variations in state
%to the width of the attenuation peak and hence to the apparent
%temperature.  Since the experiment was designed to produce as constant
%a state as practical, we assumed that variations from accelerating and
%distorted projectiles would be captured adequately through the spatial variations
%in temperature and velocity; this assumption was verified by comparing the 
%spectral attenuation calculated at different instants of time.
%It is also reasonable to assume that the natural width of the resonance and
%the time-dependence of the neutron spectrum are reproducible and therefore
%cancel with themselves when inferring a temperature.
%(The simulated attenuation spectrum involves the convolution with these
%spectra, and the inference of temperature involves the corresponding
%deconvolution.)

Given the spatial fields of mass density $\rho(\vec r)$,
temperature $T(\vec r)$, and velocity $\vec u(\vec r)$ at some instant of time
$t$,
the spectral attenuation $\alpha$ was predicted as
\begin{equation}
\alpha(E) = \int \frac{\sigma f\rho(\vec r)}A
\frac{\exp\left\{-\left[(E'-E_r)/\delta\right]^2\right\}}{\sqrt{2\pi}\delta}
\,d\vec r
\end{equation}
where $E$ is the neutron's kinetic energy,
$\sigma$ the natural cross-section,
$f$ the dopant mass fraction,
$A$ the atomic weight of the dopant,
$E_r$ the resonance energy,
\begin{equation}
E'=\frac 12 m_n |\vec u_n(E) - \vec u|^2
\end{equation}
where $m_n$ and $\vec u_n$ are the mass and velocity of the neutrons,
and
\begin{equation}
\delta = 2\sqrt{\frac{E' k_B T}A}
\label{eq:gausst}
\end{equation}
where $k_B$ is Boltzmann's constant.
\begin{equation}
\vec u_n(E) = \hat u_n\sqrt{\frac{2 E}{m_n}}
\end{equation}
where $\hat u_n$ is the direction vector of the neutron beam.
%The energy-dependent spectrum is straightforward to convert to a time-of-flight,
%but this conversion was not necessary for the subsequent analysis performed 
%here.

%As the shock reached the doped layer, the predicted spectral attenuation 
%showed the disappearance of the unshocked peak and the appearance of the
%shocked peak (Fig~\ref{fig:attenhist}).

%\subsection{Inferred temperatures}
The spectrum $\alpha(E)$
calculated at any instant of time was interpreted as an 
apparent temperature by fitting a Gaussian,
from which the apparent temperature was calculated using
Eq.~\ref{eq:gausst}.
%The self-consistency of this process was tested by comparing the
%temperature inferred for the unshocked material: 296.4\,K compared with
%293\,K as the problem was defined.
%For a simulation in which material strength was neglected, the
%inferred temperature in the shocked material was 757.3\,K.
The calculated attenuation spectrum was reproduced very well
by a Gaussian,
so it would not be possible to distinguish
spatial variations by variations in the shape of the resonance.

%\subsection{Effect of spatial variations}
%The effect of spatial variations in temperature and velocity was
%investigated by calculating spectra and inferring the temperature
%with the temperature and/or velocity set to a constant value.
%The constant value was in each case taken from a representative on-axis
%value from the simulation.
%If both fields were set constant, the simulation gave a state lying on the
%principal shock Hugoniot for Mo, as it should.
Spatial variations in temperature accounted for 24\,K of the apparent NRS temperature;
spatial variations in velocity (including radial components) accounted for
124\,K.
The contributions did not combine linearly: the combined effect was 156\,K.
The resulting apparent temperature was consistent with the Mo NRS
measurements.
The magnitude of the pressure, temperature, and contributions to the
apparent NRS temperature varied with time and position within the sample,
because of reverberations in the projectile originating with the loading
history applied at launch.
%The temperature that would be inferred in an NRS measurement depends on the
%exact synchronization of the neutron pulse with respect to the impact,
%and on the time-dependence of the source spectrum.
%As the synchronization could not be inferred {\it post facto} for the Mo
%experiments, these variations were represented as uncertainty bars in the
%simulation.
The measured NRS temperature was corrected for projectile curvature by
subtracting 156\,K.
The resulting temperature was consistent with the calculated shock Hugoniots,
and lay closer to the prediction using plastic flow
(Fig.~\ref{fig:nrs_tp}).

%\begin{figure}
%\begin{center}\includegraphics[scale=0.7]{nrs_tp5_bw.eps}\end{center}
%\caption{Apparent temperatures inferred from simulated neutron resonance data,
%   compared with the shock Hugoniot for Mo with and without strength,
%   and with simulations in which projectile curvature was included.}
%\label{fig:nrs_tp}
%\end{figure}

It can be seen that the effect of projectile flatness on the
apparent temperature should be $o(100)$\,K in these experiments 
by considering the variation in the component of material velocity 
along the neutron path inclined at $55^\circ$ to the axis for
a material speed in the shocked state of $\sim 1$\,km/s and the 
Forest Flyer projectile curvature of $\sim 5\times 10^{-3}$/mm.
The resulting shock curvature gives a speed variation of about
$\pm 120$\,m/s along the axis of the neutrons,
which equates to a effective temperature for $^{182}$W
atoms of around 105\,K when integrated around the azimuth.
This simple estimate ignores the detailed contributions from neutron speeds and
finer spatial variations which were included above,
but is of the same order as the rigorous simulation.

%\section{Conclusions}
We conclude that the NRS measurements of shock temperatures were
consistent with the published EOS and constitutive behavior of Mo,
taking into account the sensitivity of NRS to radial flow induced by
the explosively-driven projectiles used.
Plastic flow was calculated consistently with the shock jump relations.
Using published plasticity data for Mo,
plastic flow was predicted to raise the material temperature by 53\,K
compared with the hydrodynamic shock Hugoniot, in the pressure regime
of the experiments.
This goes a long way to reconciling the measurements with temperatures
expected from the various equations of state for Mo,
but does not explain the whole discrepancy.
The Forest Flyer system as described is likely to exhibit significant
spatial and temporal variations in loading as applied to the sample.
The overall effect was estimated to be around 160\,K, dominated by the
contribution from spatial variations in velocity.
The temperatures inferred from NRS seem
entirely plausible given the combined contribution of hydrodynamic shock,
plastic work, and spatial variations in loading.
The difference between the apparent temperatures was similar to the
apparent heating caused by spatial variations in the projectile.
The difference between the two points could well
reflect the finite reproducibility of the explosively-launched projectile design
used.
%A porous region in the projectile could cause partial ramp compression
%instead of a single shock, reducing the temperature.
The plastic work and non-ideal projectile contributions to apparent
temperature were all predicted using {\it a priori} modeling,
with no adjustments made to improve the match to experiment.

Although the NRS technique for measuring shock temperatures has been known
for some years, little further development has occurred because of
the temperature discrepancy in Mo.
Despite this lack of effort in development,
the per-shot uncertainty in temperature 
in the Mo experiments
was similar to that attained by mature pyrometry measurements.
There is much scope for future developments of the loading system
and the NRS diagnostic.
Flatter shocks could be induced with improved explosive launchers,
propellant guns, electromagnetic guns, or lasers.
Electromagnetic and laser loading would allow the sample to be much closer
to the neutron source, and the elimination of explosive products means smaller
neutron losses from collision with H atoms.
Other loading histories could be readily explored, including ramp
and multiple-shock compression, and release from a shock.
More recent NRS temperature measurements have incorporated a filter with
resonances bracketing those of the sample, e.g. Ag for $^{182}$W, improving
the measurement of the background during the shock experiment.
NRS measurements could be made without doping with an element of
different atomic number, possibly using the natural isotopic composition
of a material, subtracting the known resonance of unshocked material if 
necessary.
The sensitivity of the neutron detectors can be improved by optimizing the
thickness and composition of the scintillator.
Temperature uncertainties of 20-30\,K appear readily possible.
It may be possible to construct a neutron detector giving spatial
resolution.
Eventually, the resonance signal could be precise enough to allow
moments of the density of phonon states to be measured in shocked material,
as has been demonstrated statically \cite{Lynn98}.
Now that the previous NRS results are understood, the technique can be
used with more confidence for the wide range of temperature measurements
of interest in material dynamics.

%\section*{Acknowledgments}
We would like to acknowledge the contribution of
Carl Greeff for advice on equations of state for Mo, and of
Ron Rabie, David Funk, Rob Hixson, Chuck Forest, and William Buttler
for detailed information on the design, testing, and performance of the 
Forest Flyer loading system.
The work was performed under the auspices of
the U.S. Department of Energy under contracts W-7405-ENG-36
and DE-AC52-06NA25396.

\end{document}